\documentclass[aps,prb,twocolumn,preprintnumbers,amsmath,amssymb,superscriptaddress]{revtex4}%

\usepackage{graphicx}%
\usepackage{dcolumn}
\usepackage{amsmath}
\usepackage{color}

\begin{document}

\title{Magnetic tricritical point and nematicity in FeSe under pressure}

\author{Rustem~Khasanov}
 \email{rustem.khasanov@psi.ch}
 \affiliation{Laboratory for Muon Spin Spectroscopy, Paul Scherrer Institut, CH-5232 Villigen PSI, Switzerland}

\author{Rafael M. Fernandes}
 \affiliation{University of Minnesota, Minneapolis, MN 55455, USA}

\author{Gediminas Simutis}
 \affiliation{Laboratory for Muon Spin Spectroscopy, Paul Scherrer Institut, CH-5232 Villigen PSI, Switzerland}

\author{Zurab Guguchia}
 \affiliation{Laboratory for Muon Spin Spectroscopy, Paul Scherrer Institut, CH-5232 Villigen PSI, Switzerland}
 \affiliation{Department of Physics, Columbia University, New York, New York 10027, USA}

\author{Alex Amato}
 \affiliation{Laboratory for Muon Spin Spectroscopy, Paul Scherrer Institut, CH-5232 Villigen PSI, Switzerland}

\author{Hubertus Luetkens}
 \affiliation{Laboratory for Muon Spin Spectroscopy, Paul Scherrer Institut, CH-5232 Villigen PSI, Switzerland}

\author{Elvezio Morenzoni}
 \affiliation{Laboratory for Muon Spin Spectroscopy, Paul Scherrer Institut, CH-5232 Villigen PSI, Switzerland}

\author{Xiaoli Dong}
 \affiliation{Beijing National Laboratory for Condensed Matter Physics, Institute of Physics {\rm \&} University of Chinese Academy of Sciences, CAS, Beijing 100190, China}
\author{Fang Zhou}
 \affiliation{Beijing National Laboratory for Condensed Matter Physics, Institute of Physics {\rm \&} University of Chinese Academy of Sciences, CAS, Beijing 100190, China}
\author{Zhongxian~Zhao}
 \affiliation{Beijing National Laboratory for Condensed Matter Physics, Institute of Physics {\rm \&} University of Chinese Academy of Sciences, CAS, Beijing 100190, China}

\begin{abstract}
Magnetism induced by external pressure ($p$) was studied
in a FeSe crystal sample by means of muon-spin rotation. The magnetic
transition changes from second-order
to first-order for pressures exceeding the critical value $p_{{\rm c}}\simeq2.4-2.5$~GPa.
The magnetic ordering temperature ($T_{{\rm N}}$) and the value of
the magnetic moment per Fe site ($m_{{\rm Fe}}$) increase continuously
with increasing pressure, reaching $T_{{\rm N}}\simeq50$~K and $m_{{\rm Fe}}\simeq0.25$~$\mu_{{\rm B}}$
at $p\simeq2.6$~GPa, respectively. No pronounced features at both
$T_{{\rm N}}(p)$ and $m_{{\rm Fe}}(p)$ are detected at $p\simeq p_{{\rm c}}$,
thus suggesting that the stripe-type magnetic order in FeSe remains
unchanged above and below the critical pressure $p_{{\rm c}}$. A
phenomenological model for the $(p,T)$ phase diagram of FeSe reveals
that these observations are consistent with a scenario where the
nematic transitions of FeSe at low and high pressures are driven by different
mechanisms. 
\end{abstract}

\maketitle
\maketitle

\section{Introduction}

In unconventional superconductors, like heavy-fermions, cuprates and
iron-based materials, superconductivity typically emerges when
the antiferromagnetic order of the parent compound is reduced (or
fully suppressed) by changing a tuning parameter, such as doping
or pressure (see \textit{e.g.} Ref.~\onlinecite{Scalapino_RMP_2012}
for a review). The spin-density wave (SDW) antiferromagnetism in iron-based
superconductors (Fe-SC's) is, generally, of a stripe-type, \textit{i.e.},
its ordering vector points along one of the two in-plane directions.
As a consequence, magnetic order (with the ordering temperature
$T_{{\rm N}}$) becomes coupled to a tetragonal-to-orthorhombic structural
transition (with the transition temperature $T_{{\rm s}}$). Magnetism
occurs in the orthorhombic phase, whereas the paramagnetic phase can
be either tetragonal or orthorhombic. Simultaneous magnetic and structural
phase transitions ($T_{{\rm N}}=T_{{\rm s}}$) are observed, \textit{e.g.},
in Fe$_{1-y}$(Se$_{x}$Te$_{1-x}$),\cite{Li_PRB_2009} SrFe$_{2}$As$_{2}$,\cite{Jesche_PRB_2008,Wu_SciRep_2014}
(Ba$_{1-x}$K$_{x}$)Fe$_{2}$As$_{2}$,\cite{Avci_PRB_2012} and
Ba(Fe$_{1-x}$Ru$_{x}$)$_{2}$As$_{2}$.\cite{Thaler_PRB_2010} In
some Fe-SC families, like Co- or Ni-substituted BaFe$_{2}$As$_{2}$,\cite{Ni_PRB_2009,Canfield_ARoCM_2010}
LaFeAsO,\cite{DeLaCruz_Nat_2008,Luetkens_NatMat_2009} and NaFeAs,\cite{Parker_PRL_2010},
the structural transition precedes the magnetic one by several degrees
($T_{{\rm N}}<T_{{\rm s}}$). Despite the separation of $T_{{\rm s}}$
and $T_{{\rm N}}$, the two transitions are found to follow each
other rather closely as a function of tuning parameter for most Fe-SC
families, thus suggesting that the structural transition is related
to nematic electronic degrees of freedom and that the magnetic fluctuations
induce the tetragonal-to-orthorombic transition at $T_{{\rm s}}\geqslant T_{{\rm N}}$.
\cite{Fang_PRB_2008, Xu_PRB_2009, Fernandes_PRB_2012, Fernandes_PRL_2013}

FeSe, a binary pnictide belonging to a broad family of Fe-SC's,
represents an exception to the above mentioned rule. Bulk FeSe at ambient
pressure undergoes a tetragonal-to-orthorhombic transition at $T_{{\rm s}}\simeq90$~K,
\cite{Hsu_PNAS_2008,Pomjakushina_PRB_2009,McQueen_PRL_2009,Khasanov_NJP_2010}
similarly  to the nematic transition of other iron-based parent materials.
However, no magnetic order is found to occur at ambient pressure \cite{McQueen_PRL_2009,Khasanov_PRB_2008,Bendele_PRL_2010,Bendele_PRB_2012}
and FeSe superconducts below the transition temperature $T_{{\rm c}}\simeq8$~K.\cite{Hsu_PNAS_2008}
While the absence of a magnetic transition has allowed one to study
the pure nematic phase over a wide temperature range, it has also
raised the question of whether the nematicity in FeSe has the same
magnetic origin as in the other Fe-based families (see \textit{e.g.}
Ref.~\onlinecite{Bohmer_JPCM_2017} for a review). In contrast to other
Fe-SC's,\cite{Fernandes_PRL_2013} a close relationship between magnetic
and nematic fluctuations has not been observed in FeSe, thus
suggesting that other degrees of freedom may be at play.\cite{Bohmer_PRL_2015,Baek_NatMat_2015}
From the theory side, a variety of proposals were put forward to explain
the mysterious nematicity of FeSe. \cite{Valenti_NatPhys_2015, Yu_PRL_2015, DHLee_NatPhys_2015, Kontani_PRX_2016, Fanfarillo_PRB_2018, Chubukov_PRX_2016}

Properties of FeSe, however, change dramatically under applied
pressure. $T_{{\rm c}}$ rises up to a maximum value of $\simeq37$~K
at $p\simeq6$~GPa,\cite{Miyoshi_JPSJ_2012,Mizuguchi_APL_2008,Medvedev_NatMat_2009,Margadonna_PRB_2009,Garbarino_EPL_2009,Masaki_JPSJ_2009,Okabe_PRB_2010}
and a magnetically ordered phase emerges at $p\simeq0.8$~GPa.\cite{Bendele_PRL_2010,Bendele_PRB_2012}
The relation between the magnetic and structural transitions becomes
pressure dependent: while $T_{{\rm N}}$ rises continuously with increasing
pressure,\cite{Bendele_PRL_2010,Bendele_PRB_2012} $T_{{\rm s}}$
first decreases by reaching $T_{{\rm s}}\simeq20$~K at $p\simeq1.6$~GPa,\cite{Miyoshi_JPSJ_2012}
and then increases again by approaching $T_{{\rm s}}\simeq30$~K
at $p\simeq3$~GPa.\cite{Kothapalli_NatComm_2016,Bohmer_Arxiv_2018}
It is worth to emphasize, that for $p\lesssim1.6$~GPa the appearance of magnetism for $p\gtrsim0.8$~GPa
has little influence on the $T_{{\rm s}}(p)$ phase boundary thus pointing
to an independence of the magnetic and structural transitions in this
pressure range. Interestingly, the high-pressure behavior
of FeSe resembles the situation observed in other Fe-SC's. Indeed,
above $p\simeq1.6$~GPa the structural and magnetic transitions follow
each other,\cite{Kothapalli_NatComm_2016} and they merge into a combined
first-order like transition for pressures exceeding $\simeq2.2$~GPa.\cite{Wang_PRL_2016}

It is quite likely, therefore, that for FeSe the external pressure
plays the role of a tuning parameter that changes the driving force
of nematicity from a yet to be determined mechanism at low $p$ to
the usual magnetic mechanism of other Fe-SC's for pressures exceeding
a certain critical value $p_{\rm c}$. In order to check the validity of such assumption,
muon-spin rotation ($\mu$SR) experiments under pressures up to $p\simeq2.64$~GPa
on a FeSe crystal sample were performed. The results obtained in the
present study suggest that the magnetic transition in FeSe changes
from second-order for $p\lesssim2.4$~GPa to first-order for pressures
exceeding $p_{{\rm c}}\simeq2.4-2.5$~GPa, thus signaling for the occurrence
of a magnetic tricritical point. This observation is explained via a
phenomenological Ginzburg-Landau model where the nematic transition
at low pressures ($p<p_{c}$) has a different origin than the magnetically-driven vestigial
nematicity at higher pressures ($p>p_{c}$). While other scenarios
may also be compatible with this phenomenological model,\cite{Valenti_NatPhys_2015,Yu_PRL_2015,DHLee_NatPhys_2015,Kontani_PRX_2016,Fanfarillo_PRB_2018}
our findings are consistent with the mechanism proposed in Ref.~\onlinecite{Chubukov_PRX_2016}.
The renormalization-group calculations presented in that work reveal that the small value of the Fermi energy (which is the case for FeSe at ambient,\cite{Kasahara_PNAS_2014, Terashima_PRB_2014, Watson_PRB_2015, Audouard_EPL_2015} and at low pressures) makes
a $d-$wave Pomeranchuk transition the leading instability of the
system. The magnetism in this case remains weak.  Upon increasing the value of
the Fermi energy (which for FeSe is presumably accomplished by increasing the pressure) the
magnetism becomes the leading instability, and the Pomeranchuk instability
is suppressed. In this regime, the nematicity can only arise as a vestigial
phase of the magnetically ordered state.

The paper is organized as follows. Sections~\ref{subsec:Sample-Preparation}
and \ref{subsec:Experimental-Techniques} describe the sample preparation
procedure and experimental techniques. The results obtained in the
zero-field and weak transverse-field $\mu$SR experiments are summarized
in Secs.~\ref{subsec:ZF-muSR} and \ref{subsec:wTF-muSR}. Section~\ref{sec:Theoretical-model}
presents the theoretical phenomenological model describing the emergence
of the magnetic tricritical point due to coupled Pomeranchuk and SDW
magnetic instabilities. In Section~\ref{sec:Discussions} the dependence
of the ordered moment on the magnetic ordering temperature (Sec.~\ref{subsec:Moment-on-Tn}),
the dependence of $T_{{\rm N}}$ on pressure (Sec.~\ref{subsec:Tn-on-p})
and the consistency of the second- and the first-order type transitions
with the theory (Sec.~\ref{subsec:Consistency-Second-First}) are
discussed. The conclusions follow in Sec.~\ref{sec:Conclusions}.

\section{Experimental details}

\label{sec:Experiment}

\subsection{Sample preparation}

\label{subsec:Sample-Preparation}

The FeSe crystal was synthesized by means of floating zone technique
as described in Ref.~\onlinecite{Ma_SST_2014}. X-ray measurements
confirm that the grown cylindrical sample exhibits a single preferred
orientation of a tetragonal (101) plane.\cite{Ma_SST_2014}

\subsection{Experimental Techniques}

\label{subsec:Experimental-Techniques}

\subsubsection{Pressure Cell}

The pressure was generated in a double-wall piston-cylinder type of
cell made of MP35N alloy. As a pressure transmitting medium 7373 Daphne
oil was used. This oil solidifies at $p\simeq2.3$~GPa at room temperature,\cite{Murata_RSI_1997}
meaning that the experiments for $p<2.3$~GPa were conducted in hydrostatic
conditions, while for higher pressures the conditions were quasi-hydrostatic.

The pressure was measured in situ by monitoring the pressure iduced shift
of the superconducting transition temperature of In (pressure indicator).
The details of the experimental setup for conducting $\mu$SR under
pressure experiments are given in Refs.~\onlinecite{Khasanov_HPR_2016}
and \onlinecite{Shermadini_HPR_2017}.

\subsubsection{Muon-spin rotation}

The zero-field (ZF) and weak transverse-field (wTF) muon-spin rotation
($\mu$SR) experiments were carried out at the $\mu$E1 beam line
by using the GPD (General Purpose Decay) spectrometer (Paul Scherrer Institute,
Switzerland).\cite{Khasanov_HPR_2016} Measurements
were performed for temperatures ranging from $T\simeq3$ to $\simeq120$~K
and pressures in the range of $1.7\leq p\leq2.64$~GPa. The experimental
data were analyzed by using the MUSRFIT package.\cite{Suter_MUSRFIT_2012}

The $\mu$SR data were analyzed by decomposing the signal on the sample
(s) and the pressure cell (pc) contributions:
\begin{equation}
A(t)=A_{{\rm s}}(0)P_{{\rm s}}(t)+A_{{\rm pc}}(0)P_{{\rm pc}}(t),\label{eq:ZF_Asymmetry_PC}
\end{equation}
Here $A_{{\rm s}}(0)$ and $A_{{\rm pc}}(0)$ are the initial asymmetries
and $P_{{\rm s}}(t)$ and $P_{{\rm pc}}(t)$ are the muon-spin polarizations
belonging to the sample and the pressure cell, respectively. The polarization
of the pressure cell $P_{{\rm pc}}(t)$ was obtained in a separated
set of experiments.\cite{Khasanov_HPR_2016} In the data analysis
the ratio of the component of the pressure cell and the component
of the sample $A_{{\rm s}}(0)/A_{{\rm pc}}(0)$ was kept constant
for each individual pressure and was always $\approx80$\%.

The analysis of the ZF-$\mu$SR response of the FeSe sample was made
by considering that the magnetic order appears gradually in volume.\cite{Bendele_PRL_2010,Bendele_PRB_2012}
One part of the muons experiences a static local field corresponding
to the magnetic order and the other part stops in nonmagnetic regions:
\begin{eqnarray}
P_{{\rm s}}^{{\rm ZF}}(t) & = & m^{{\rm ZF}}\left[f_{{\rm osc}}e^{-\lambda_{T}t}\cos(\gamma_{\mu}B_{{\rm int}}t)+(1-f_{{\rm osc}})\;e^{-\lambda_{L}t}\right]\nonumber \\
 &  & +(1-m^{{\rm ZF}})\;e^{-\lambda_{0}^{{\rm ZF}}t}.\label{eq:P_AFM}
\end{eqnarray}
Here $m^{{\rm ZF}}$ is the magnetic volume fraction of the sample,
$B_{{\rm int}}$ is the internal field on the muon stopping site,
$\gamma_{\mu}=2\pi\cdot135.5$~MHz/T is the muon gyromagnetic ratio,
and $\lambda_{T}$ and $\lambda_{L}$ are the transverse and the longitudinal
exponential relaxation rates, respectively. $\lambda_{0}^{{\rm ZF}}$
is the exponential rate in the non-magnetic parts of the sample. The
oscillating ($f_{{\rm osc}}$) and non-oscillating ($1-f_{{\rm osc}}$)
fractions arise from muons sensing the internal field components which are transversal {[}$B_{{\rm int}}\perp P(0)${]}
and longitudinal {[}$B_{{\rm int}}\parallel P(0)${]} to the initial
muon-spin polarization, respectively. Note that since the FeSe crystal
sample studied here had one preferable orientation (101 orientation,
see Refs.~\onlinecite{Ma_SST_2014,Khasanov_PRB_2017}) the value of
$f_{{\rm osc}}\simeq0.75$ was different from that expected for a
polycrystalline sample ($f_{{\rm osc}}\equiv2/3$), where all angles between
$B_{{\rm int}}$ and $P(0)$ are equally possible.\cite{Yaouanc_book_2011}

The wTF-$\mu$SR sample response was analyzed considering that the
muons stopping in a non-magnetic environment produce long lived oscillations,
which reflect the coherent muon-spin precession around the external field
$B_{{\rm ex}}$.
\begin{equation}
P_{{\rm s}}^{{\rm wTF}}(t)=(1-m^{{\rm wTF}})e^{-\lambda_{0}^{{\rm wTF}}t}\;\cos(\gamma_{\mu}B_{{\rm ex}}t+\phi).\label{eq:wTF}
\end{equation}
Here $m^{{\rm wTF}}$ is the magnetic volume fraction of the sample,
$\phi$ is the initial phase of the muon-spin ensemble and $\lambda_{0}^{{\rm wTF}}$
is the exponential depolarization rate. Note that within the weak
transverse-field regime ($B_{{\rm ex}}\ll B_{{\rm int}}$) and for
the short lived oscillations of the muon-spin polarization in magnetically
ordered parts of the sample [as is the case for FeSe, see {\it e.g.} Fig.~\ref{fig:time-specra_Fourier}~(a)] one neglects the magnetic contribution.

\section{Experimental results }

\subsection{Zero-field $\mu$SR experiments}
\label{subsec:ZF-muSR}

\begin{figure}[tb]
\includegraphics[width=1\linewidth]{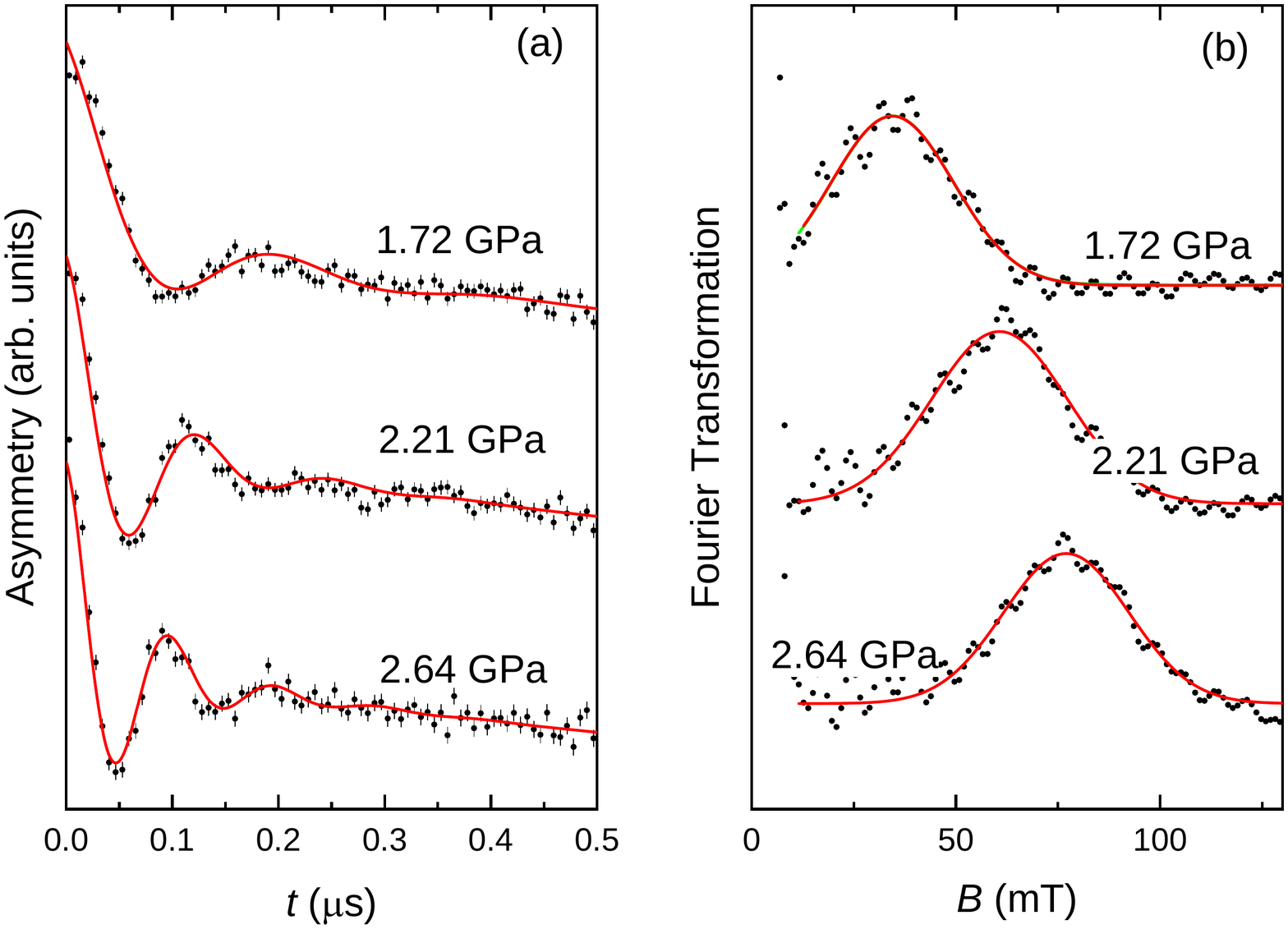} 
\caption{(a) The muon time spectra collected at pressures $p\simeq1.72$, 2.21
and 2.64~GPa. The red lines are fits of Eq.~\ref{eq:ZF_Asymmetry_PC},
with the sample contribution described by Eq.~\ref{eq:P_AFM} to
the experimental data. In order to increase the counting statistics
the $\mu$SR spectra accumulated in the
temperature range from $\simeq3$ to 20~K were summed together. (b)
The Fourier transform of the ZF-$\mu$SR data shown in the panel (a). The
red lines are Lorentzian fits.}
\label{fig:time-specra_Fourier}
\end{figure}

Figure~\ref{fig:time-specra_Fourier}~(a) shows the muon-time spectra
at pressures $p\simeq1.72$, 2.21 and 2.64~GPa. In order to increase
the counting statistics (to decrease the error bars) the $\mu$SR
spectra accumulated in the temperature range from $\simeq3$ to 20~K
were added. The red lines correspond to the fit of Eq.~\ref{eq:ZF_Asymmetry_PC}
with the sample contribution described by Eq.~\ref{eq:P_AFM} to
the experimental data. The spontaneous muon-spin precession reflects
the appearance of a static magnetic order below the N\'{e}el temperature
$T_{{\rm N}}$. The field distributions obtained by Fourier transform
of the ZF-$\mu$SR spectra are shown in Fig.~\ref{fig:time-specra_Fourier}~(b).

From the data presented in Fig.~\ref{fig:time-specra_Fourier} the
following points can be concluded: (i) The field distribution in the FeSe sample
is well described by a single  Lorentzian (see Eq.~\ref{eq:P_AFM}). This suggests
that the magnetic order is {\it commensurate},
which is consistent with earlier $\mu$SR measurements on FeSe
polycrystalline samples. \cite{Bendele_PRL_2010,Bendele_PRB_2012}
(ii) The width of the field distribution is almost pressure independent
($\Delta B_{{\rm int}}\simeq30$~mT). Bearing in mind that the internal
field increases with increasing pressure ($B_{{\rm int}}\simeq34$,
61, 77~mT for $p=1.72$, 2.21, and 2.64~GPa, respectively), this
would imply that the magnetic field distribution becomes more homogenous.
The distribution of internal fields, and the corresponding ordered magnetic moments
per Fe site ($m_{{\rm Fe}}\propto B_{{\rm int}}$, see \textit{e.g.}
Ref.~\onlinecite{Yaouanc_book_2011}), have values $\Delta B_{{\rm int}}/B_{{\rm int}}=\Delta m_{{\rm Fe}}/m_{{\rm Fe}}\simeq45$\%,
25\%, and 19\% for $p=1.72$, 2.21, and 2.64~GPa, respectively. (iii)
The increase of the internal field is caused by the corresponding
increase of the ordered magnetic moments. Following Ref.~\onlinecite{Khasanov_PRB_2017},
where for the stripe-type magnetic order of FeSe the value of $B_{{\rm int}}=0.31-0.32$~T
per 1~$\mu_{{\rm B}}$ per Fe atom was determined, the ordered magnetic
moment is estimated to be $m_{{\rm Fe}}\simeq0.11$, $0.19$, and
0.25~$\mu_{{\rm B}}$ for $p=1.72$, 2.21, and 2.64~GPa, respectively.

\begin{figure}[htb]
\includegraphics[width=1\linewidth]{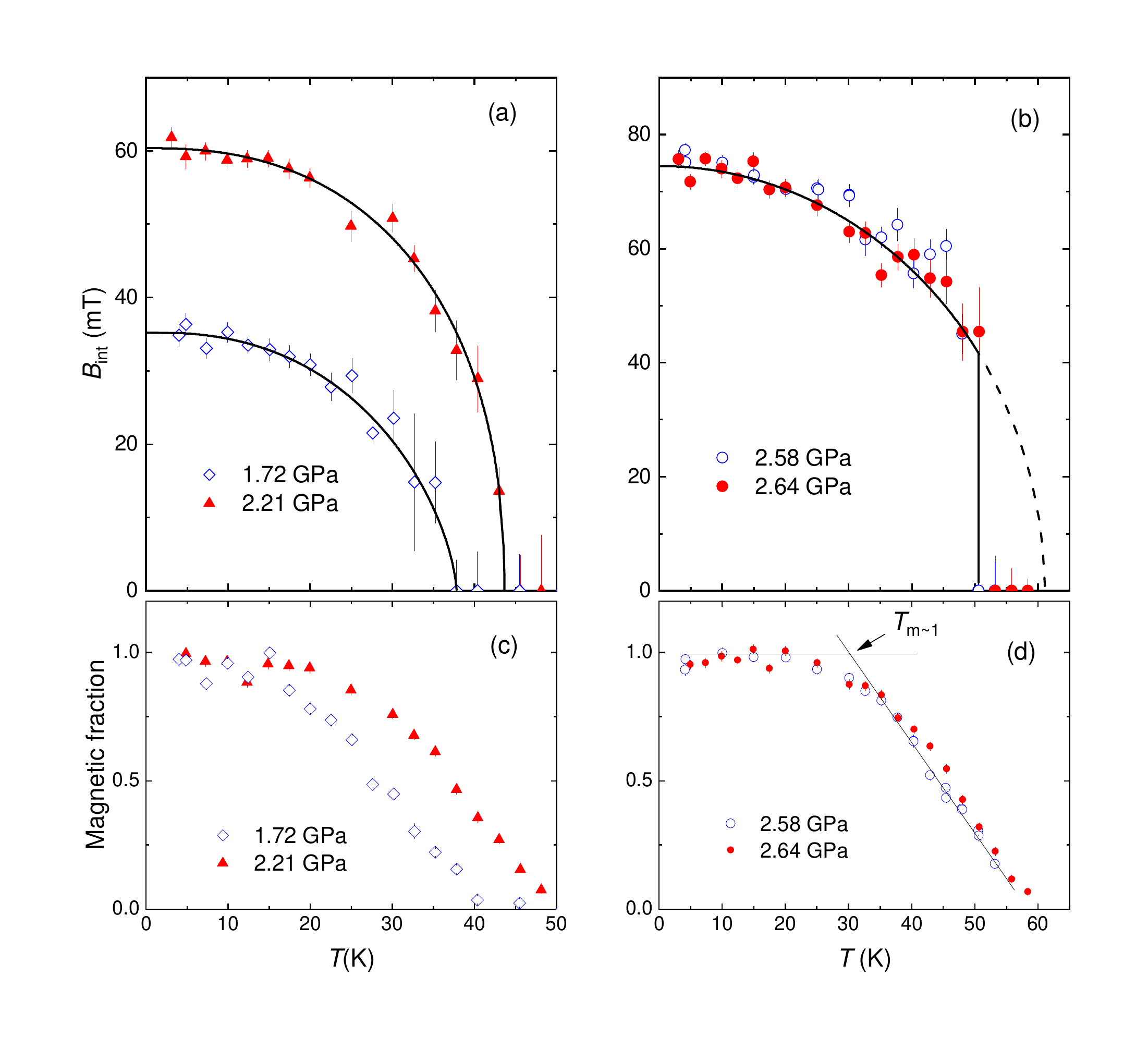} 
\caption{(a) Temperature dependence of the internal field $B_{{\rm int}}$
at $p=1.72$ and 2.21~GPa. (b) The same as in panel (a) but for $p=2.58$
and 2.64~GPa. The lines in (a) and (b) are fits of Eq.~\ref{eq:Power-Law}
to $B_{{\rm int}}(T)$ data. (c) Temperature dependence of the magnetic
volume fraction $m^{{\rm ZF}}$ at $p=1.72$ and 2.21~GPa. (d) The
same as in panel (c) but for $p=2.58$ and 2.64~GPa. The temperature
$T_{{\rm m\sim1}}$ (the temperature where the magnetic volume fraction
starts to deviate from unity) is determined as a crossing point of
$m^{{\rm ZF}}=1$ line with the linear fit of $m^{{\rm ZF}}(T)$ in
the vicinity of the magnetic transition {[}panel (d){]}. }
\label{fig:Bint_Fraction}
\end{figure}

Figure~\ref{fig:Bint_Fraction} shows the temperature dependence of
the internal field $B_{{\rm int}}$ and of the magnetic volume fraction
$m^{{\rm ZF}}$ obtained from the fit of ZF-$\mu$SR data, for a few
characteristic pressures. Obviously, the $B_{{\rm int}}(T)$ dependencies
at ``low'' ($p\simeq1.72$ and 2.21~GPa) and ``high'' ($p\simeq2.58$
and 2.64~GPa) pressures are quite different. At low pressures
{[}Fig.~\ref{fig:Bint_Fraction}~(a){]}, $B_{{\rm int}}$ appears
to decrease continuously with increasing temperature until it vanishes
at $T_{{\rm N}}$. This behavior is typical for a second-order transition,
which is characterized by a continuous decrease of the order parameter
by approaching the critical temperature. In contrast, at higher pressures
{[}Fig.~\ref{fig:Bint_Fraction}~(b){]} $B_{{\rm int}}$ drops abruptly
for temperatures slightly above 50 and 52~K for $p=2.58$ and 2.64~GPa,
respectively. This suggests that the transition becomes first-order.
Indeed, the fit of the high-pressure data up to $T\simeq 50$~K by means of a power law:
\begin{equation}
B_{{\rm int}}(T)=B_{{\rm int}}(0)\;\left[1-(T/T_{{\rm N}})^{\alpha}\right]^{\beta}\label{eq:Power-Law}
\end{equation}
($\alpha$ and $\beta$ are the power exponents) suggests that a smooth vanishing of the order parameter in a second order phase transition could be expected around 60~K. Instead an abrupt first order like transition is observed at $T\simeq 50$~K.

\begin{figure}[htb]
\includegraphics[width=0.7\linewidth]{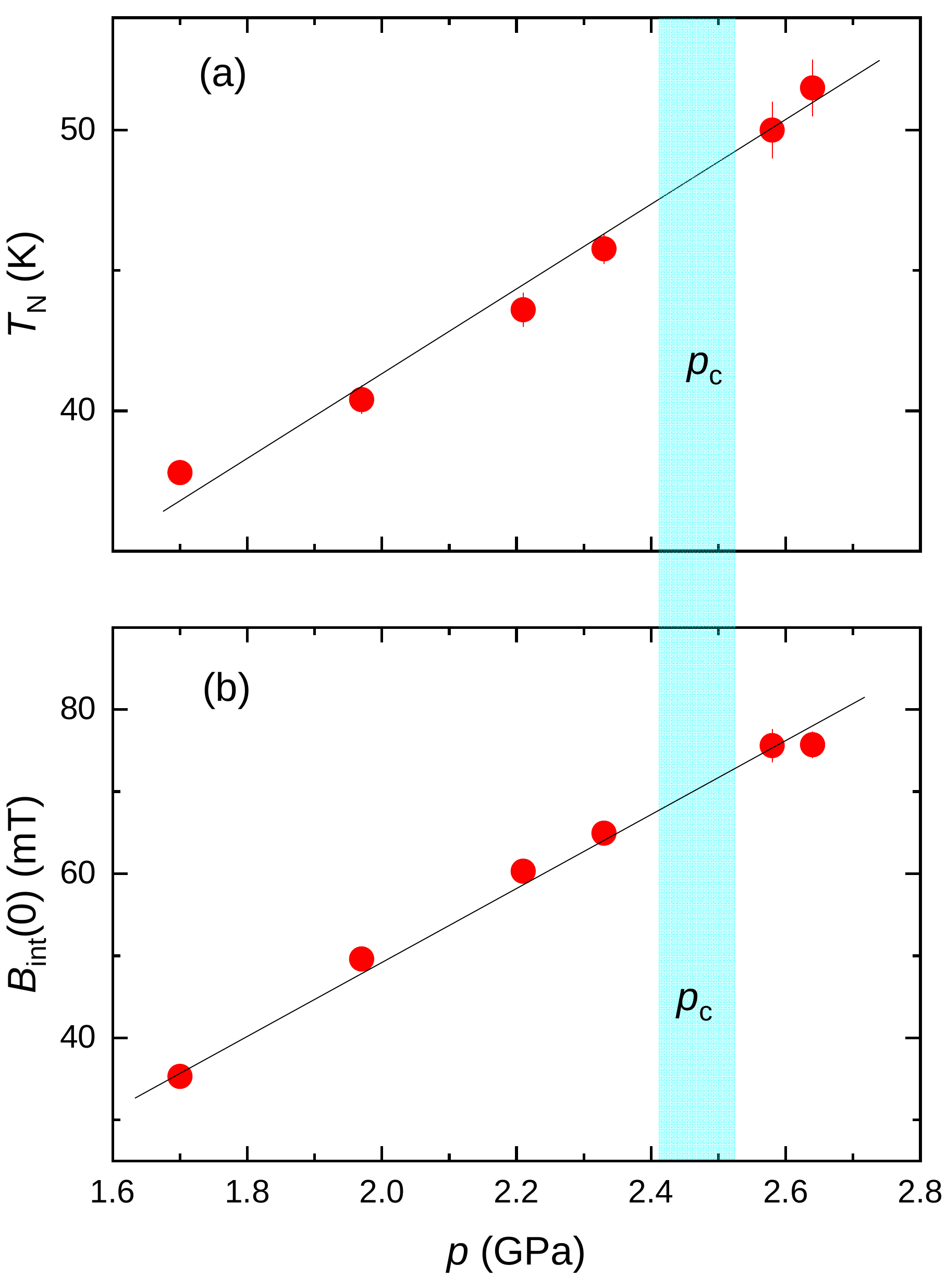} 
\caption{(a) Pressure dependence of the magnetic ordering temperature $T_{{\rm N}}$
obtained from the fit of Eq.~\ref{eq:Power-Law} to the experimental
$B_{{\rm int}}(T)$ data. The $T_{{\rm N}}$ points for $p=2.58$
and 2.64~GPa correspond to temperatures where $B_{{\rm int}}(T)$
drops to zero {[}see Fig.~\ref{fig:Bint_Fraction}~(b){]}. (b) Pressure
dependence of the zero-temperature value of the internal field. The
blue stripe represents the pressure region where the magnetic transition
changes from second-order to first-order, {\it i.e.} where a magnetic tricritical
point exists. }
\label{fig:Bint_Tn}
\end{figure}

The values of the magnetic ordering temperature $T_{{\rm N}}$ (except
for $p=2.58$ and 2.64~GPa) and the zero-temperature values of the
internal field $B_{{\rm int}}(0)$ obtained from the fit of $B_{{\rm int}}(T)$
by using Eq.~\ref{eq:Power-Law} are plotted in Fig.~\ref{fig:Bint_Tn}.
The $T_{{\rm N}}$ points for $p=2.58$ and 2.64~GPa correspond to
temperatures where $B_{{\rm int}}(T)$ drops to zero {[}see Fig.~\ref{fig:Bint_Fraction}~(b){]}.
Figure~\ref{fig:Bint_Tn} implies that both $T_{{\rm N}}$ and $B_{{\rm int}}(0)$
increases linearly with increasing temperature just following the
tendency observed in earlier $\mu$SR experiments on polycrystalline
FeSe samples.\cite{Bendele_PRL_2010,Bendele_PRB_2012} The blue stripe
corresponds to the critical pressure $p_{{\rm c}}$ where the magnetic
transition changes from second-order to first-order. It is important
to note that $T_{{\rm N}}$ and $B_{{\rm int}}(0)$ go smoothly through
$p_{{\rm c}}$ without showing any pronounced features. This indicates that the type of the magnetic order, namely stripe-type
magnetism, stays the same above and below the critical pressure $p_{c}$.

Our experiments are consistent with other measurements where a change
from second-order to first-order transition was observed.\cite{Wang_PRL_2016,Kothapalli_NatComm_2016}
Interestingly, the exact values of critical pressure $p_{c}$ seem
to vary between the different experiments. One of the possible reasons
for the discrepancy could be a sample-dependence. Also, since the measurements
were done in different pressure cells, the degree of hydrostaticity
is likely to vary from experiment to experiment. Therefore
an alternative explanation is that the critical pressure may depend
on the exact stresses in the sample.

\subsection{Weak transverse-field $\mu$SR experiments}

\label{subsec:wTF-muSR}

``Supercooling'' and ``superheating'' across a first-order
transition yield metastable states, resulting in hysteresis. In order
to search for a possible hysteretic behavior of the magnetic transition
in the FeSe crystal sample studied here, wTF-$\mu$SR experiments
were performed for pressures below (2.05~GPa) and above (2.58~GPa)
the critical pressure $p_{{\rm c}}\sim2.4-2.5$~GPa (see Fig.~\ref{fig:Bint_Tn}).
Note that the $\mu$SR experiments under weak transverse-field applied
perpendicular to the muon-spin polarization are a straightforward
method to determine the onset of the magnetic transition and the magnetic
volume fraction. In this case the contribution to the asymmetry from
muons experiencing a vanishing internal spontaneous magnetization
can be accurately determined.\cite{Khasanov_ScRep_2015}

\begin{figure}[htb]
\includegraphics[width=1\linewidth]{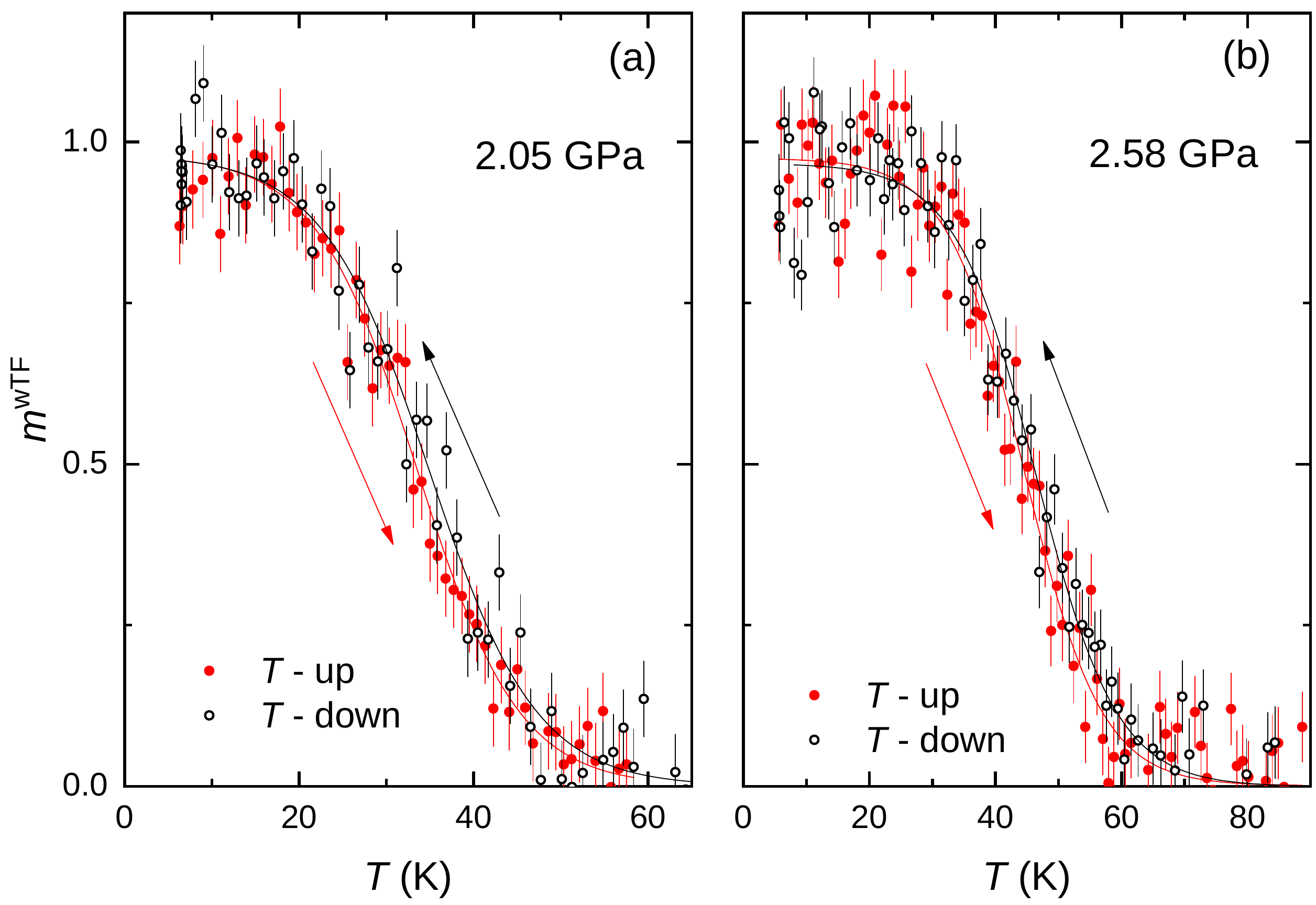} 
\caption{(a) Temperature evolution of the magnetic volume fraction $m^{{\rm wTF}}$
of FeSe obtained in the wTF-$\mu$SR measurements at $p=2.05$~GPa.
Closed and open symbols correspond to the experimental data obtained
with increasing and decreasing temperature (the sweeping rate is $\simeq0.2$~K/min,
5 minutes per data point). (b) The same as in (a) but for $p=2.58$~GPa. }
\label{fig:WTF}
\end{figure}

The temperature dependencies of the magnetic volume fraction $m^{{\rm wTF}}$
obtained from fits of Eq.~\ref{eq:ZF_Asymmetry_PC} with the sample
contribution described by Eq.~\ref{eq:wTF} to the wTF-$\mu$SR data
are summarized in Fig.~\ref{fig:WTF}. The cooling/warming rates
were set to 0.2~K/min. During the warming/cooling process the wTF-$\mu$SR
spectra were accumulated continuously (5 mins per data point). The
solid lines in Fig.~\ref{fig:WTF} correspond to fits of the equation:\cite{Khasanov_PRL_2008}
\begin{equation}
m^{{\rm wTF}}(T)/m^{{\rm wTF}}(0)=a (1+\exp[(T-T_{{\rm N}})/\Delta T_{{\rm N}}])^{-1}.\label{eq:wTF-Tn}
\end{equation}
Here $m^{{\rm wTF}}(0)$ is the magnetic volume fraction at zero-temperature,
$\Delta T_{{\rm N}}$ is the width of the magnetic transition, and $a$ is an adjusting coefficient.
The analysis reveals the presence of a small but measurable hysteresis
for both pressures. The corresponding $T_{{\rm N}}^{{\rm up}}/T_{{\rm N}}^{{\rm down}}$
values are 33.6(3)/34.9(5)~K and 44.6(6)/46.4(6)~K for $p=2.05$
and 2.58~GPa, respectively. This implies that the shift of the magnetic
ordering temperatures [$T_{{\rm N}}^{{\rm up}}-T_{{\rm N}}^{{\rm down}}=1.3(8)$
for $p=2.05$~GPa and 1.8(8)~K for $p=2.58$~GPa] is the same (within
the experimental uncertainties) above and below the
critical pressure $p_{{\rm c}}$. Bearing in mind that the type of
magnetic transition appears to change by crossing $p_{c}$ (from the second-
to the first-order type, see Fig.~\ref{fig:Bint_Fraction}), the
similar $T_{{\rm N}}^{{\rm up}}-T_{{\rm N}}^{{\rm down}}$ values
suggest that the hysteresis observed in our experiments is purely
instrumental and is probably caused by difference in thermalization
of the pressure cell during warming/cooling procedure.

Note that an approximate $1.5$~K hysteresis shift was observed by
Wang \textit{et al.}\cite{Wang_PRL_2016} in NMR experiments. Such a
temperature shift should be measurable within our experimental accuracy.
Further measurements are needed to clarify the reason for the absence of hysteresis in our wTF-experiments.

\section{Theoretical model}

\label{sec:Theoretical-model}

In this section, we use a general phenomenological model to show that
the experimentally observed emergence of the magnetic tricritical
point with pressure is consistent with a scenario in which nematicity
is driven by different mechanisms at low pressures and at high pressures.
Let us denote the nematic order parameter at low pressures by $\eta$.
For our analysis, while the specific microscopic origin of $\eta$
is not important, the main point is that it does not arise from the
usual Ising-nematic vestigial phase associated with partially melted
$\mathbf{Q}_{1}=\left(\pi,0\right)$ and $\mathbf{Q}_{2}=\left(0,\pi\right)$
stripe spin density-wave (SDW). The SDW order parameters associated
with these ordering wave-vectors are denoted by $\mathbf{M}_{1}$
and $\mathbf{M}_{2}$. Hereafter, to distinguish $\eta$ from the
magnetic-driven nematic order parameter $M_{1}^{2}-M_{2}^{2}$, we
will refer to the former as the Pomeranchuk order parameter. The Ginzburg-Landau
(GL) free energy of the coupled order parameters is given by (see
for instance Refs.~\onlinecite{Fernandes_PRB_2012,Chubukov_PRX_2016}):

\begin{align}
F\left[\eta,M_{i}\right] & =\frac{a_{p}}{2}\eta^{2}+\frac{u_{p}}{4}\eta^{4}+\frac{a_{m}}{2}\left(M_{1}^{2}+M_{2}^{2}\right)\nonumber \\
 & +\frac{u_{m}}{4}\left(M_{1}^{2}+M_{2}^{2}\right)^{2}-\frac{g_{m}}{4}\left(M_{1}^{2}-M_{2}^{2}\right)^{2}\nonumber \\
 & -\lambda\eta\left(M_{1}^{2}-M_{2}^{2}\right)\label{F}
\end{align}

Here, $a_{p}$, $u_{p}$ are the Pomeranchuk GL parameters; $a_{m}$,
$u_{m}$, $g_{m}$ are the SDW GL parameters; and $\lambda$ is the
coupling constant. To mimic the experimental situation, we assume
that pressure suppresses the Pomeranchuk transition and at the same
time enhances the SDW transition. This can be modeled by setting $a_{p}=a_{p,0}\left(T-T_{0}-\delta\right)$
and $a_{m}=a_{m,0}\left(T-T_{0}+\delta\right)$, where $a_{p,0}$
and $a_{m,0}$ are positive prefactors, and $T_{0}$ is the mean-field
transition temperature in which the Pomeranchuk and magnetic transitions
meet. The parameter $\delta$, assumed here to be only pressure dependent,
selects the leading instability to be either the Pomeranchuk transition
($\delta>0$) or the SDW transition ($\delta<0$). We therefore define
the pressure $p^{*}$ in which the two transitions meet by setting
$\delta\left(p^{*}\right)=0$, implying $\delta\propto p^{*}-p$.
Note that, while we assumed a symmetric change in the transition temperatures
with respect to $T_{0}$, the results derived here are more general.

\begin{figure}[tb]
\centering{}\includegraphics[width=0.8\linewidth]{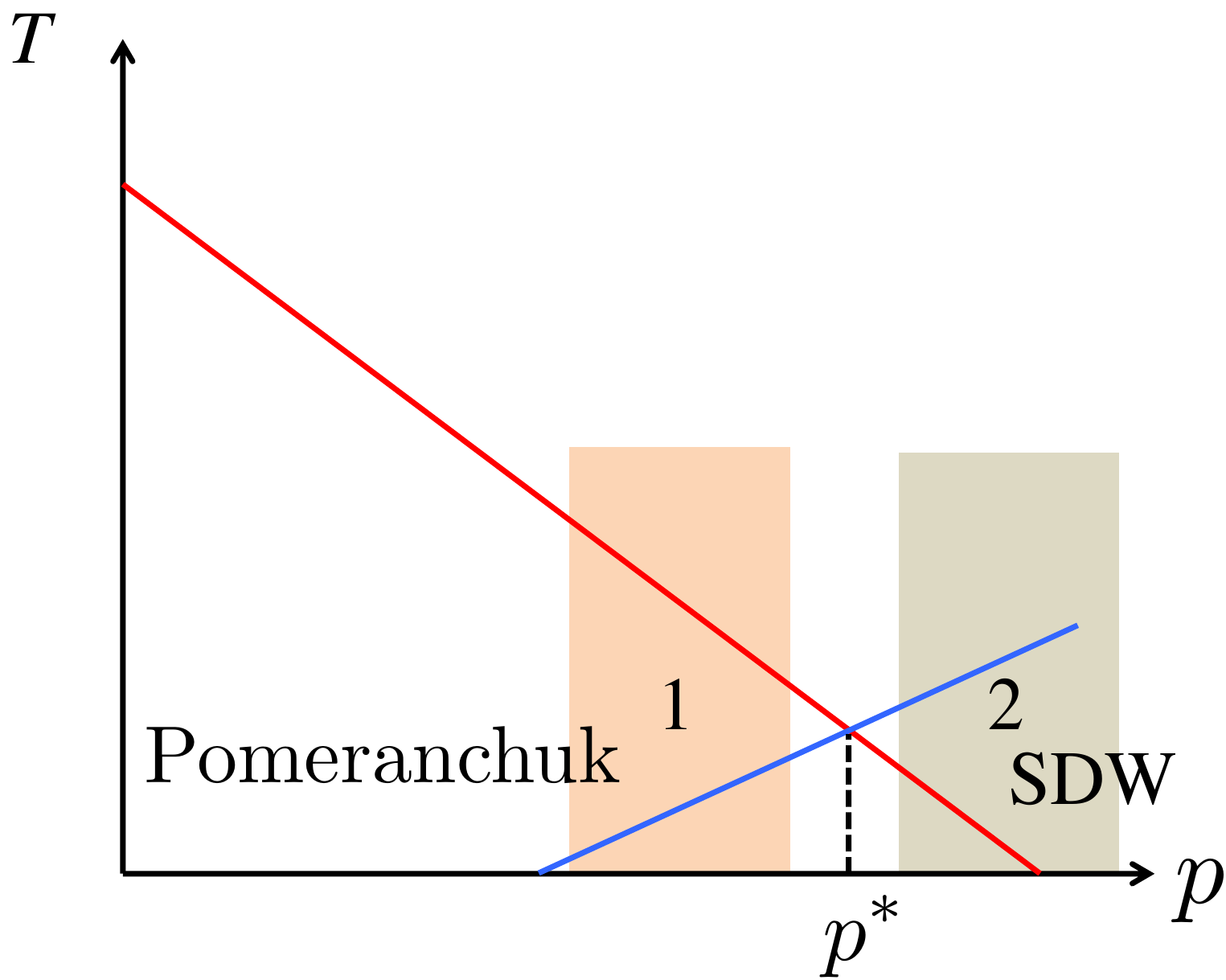}
\caption{Schematic $(p,T)$ phase diagram of FeSe used in the phenomenological
calculation. The Pomeranchuk and magnetic (SDW) transitions meet at
the multi-critical point at $\left(p^{*},T_{0}\right)$. The respective
transition temperatures are thus modeled as $T_{p}=T_{0}+\delta$
and $T_{m}=T_{0}-\delta$, with $\delta\propto p^{*}-p$ such that
$\delta>0$ refers to the Pomeranchuk side and $\delta<0$ refers
to the SDW side.}
\label{fig:phase-diagram}
\end{figure}

We can now investigate the character of the magnetic transition as
$\delta$ changes. As shown in Fig. \ref{fig:phase-diagram}, we consider
two regions in the regime $p<p^{*}$ (region 1) and $p>p^{*}$ (region
2). In region 1, the Pomeranchuk transition happens first at $T_{p}=T_{0}+\delta$.
Therefore, we first minimize the GL free energy with respect to the
Pomeranchuk order parameter $\eta$

\begin{equation}
a_{p}\eta+u_{p}\eta^{3}=\lambda\left(M_{1}^{2}-M_{2}^{2}\right)
\end{equation}

Expanding around the bare Pomeranchuk order parameter $\eta_{0}=\sqrt{-\frac{a_{p}}{u_{p}}}$
up to quartic order in $M_{1,2}$ and substituting back in Eq. (\ref{F}),
we find the effective SDW free energy:

\begin{align}
 & \tilde{F}\left[M_{i}\right]=\frac{a_{m}}{2}\left(M_{1}^{2}+M_{2}^{2}\right)-\lambda\eta_{0}\left(M_{1}^{2}-M_{2}^{2}\right)\label{F_region1}\\
 & +\frac{u_{m}}{4}\left(M_{1}^{2}+M_{2}^{2}\right)^{2}-\frac{1}{4}\left[g_{m}+\frac{\lambda^{2}}{\left(-a_{p}\right)}\right]\left(M_{1}^{2}-M_{2}^{2}\right)^{2}\nonumber
\end{align}

Thus, the onset of Pomeranchuk order has two effects on the SDW degrees
of freedom. The first one, arising from the quadratic coefficients,
is to select $M_{1}$ over $M_{2}$ (since we chose $\eta_{0}>0$).
This enhances the magnetic transition temperature $T_{m}$ from $T_{m}^{(0)}=T_{0}-\delta$
to:

\begin{equation}
T_{m}\approx T_{m}^{(0)}+2\delta-\delta^{2}\left(\frac{a_{m,0}^{2}u_{p}}{\lambda^{2}a_{p,0}}\right)=T_{p}-\delta^{2}\left(\frac{a_{m,0}^{2}u_{p}}{\lambda^{2}a_{p,0}}\right)
\end{equation}
where we expanded to leading orders in $\delta$. The second effect
is to suppress the effective quartic coefficient. Setting $M_{2}=0$,
the effective quartic coefficient is, for small but finite $\delta$,

\begin{equation}
\tilde{u}_{m}\approx u_{m}-g_{m}-\frac{1}{\delta^{2}}\left(\frac{\lambda^{4}}{a_{m,0}^{2}u_{p}}\right)
\end{equation}

Thus, for sufficiently small $\delta$, the magnetic transition becomes
first order, since $\tilde{u}_{m}$ becomes negative. This signals
the onset of a tricritical point at a pressure $p_{c}$ slightly below
$p^{*}$.

To proceed, we now show that the SDW transition remains first-order
for $p>p^{*}$, corresponding to $\delta<0$ (region 2 of Fig. \ref{fig:phase-diagram}).
In this region, because there is no long-range Pomeranchuk order above
the magnetic transition, we can use a Gaussian approximation for the
Pomeranchuk free energy (as long as $\left|\delta\right|\neq0$).
In this case, the Pomeranchuk GL equation is straightforward:
\begin{equation}
\eta=\frac{\lambda}{a_{p}}\left(M_{1}^{2}-M_{2}^{2}\right)
\end{equation}

Substituting this in Eq. (\ref{F}), we find the effective magnetic
free energy:

\begin{eqnarray}
 &  & \tilde{F}\left[M_{i}\right]=\frac{a_{m}}{2}\left(M_{1}^{2}+M_{2}^{2}\right)+\frac{u_{m}}{4}\left(M_{1}^{2}+M_{2}^{2}\right)^{2}\nonumber \\
 &  & -\frac{1}{4}\left(g_{m}+\frac{2\lambda^{2}}{a_{p}}\right)\left(M_{1}^{2}-M_{2}^{2}\right)^{2}
\end{eqnarray}

Note that, in contrast to Eq. (\ref{F_region1}), the magnetic transition
is not affected. The only term affected is the quartic coefficient
$g_{m}$, which becomes, for small but finite $\delta<0$:

\begin{equation}
\tilde{g}_{m}=g_{m}+\frac{\lambda^{2}}{a_{p,0}\left(-\delta\right)}
\end{equation}

Clearly, $g_{m}$ gets a large enhancement as the multi-critical point
is approached. The key point is that the nematic transition in this
regime arises as a vestigial order of the magnetic state, via the
condensation of the composite order parameter $\left(M_{1}^{2}-M_{2}^{2}\right)$.
This effect can only be captured beyond mean-field. According to the
large-$N$ results of Ref.~\onlinecite{Fernandes_PRB_2012}, the vestigial
Ising-nematic transition and the primary magnetic transitions are
simultaneous and first-order for large enough $\tilde{g}_{m}$ (even
if it is still smaller than $u_{m}$). In particular, for an anisotropic
3D system with effective dimensionality $2<d<3$, the simultaneous
first-order transition takes place for $\tilde{g}_{m}>\left(3-d\right)u_{m}$,
which implies, in terms of $\delta$ (recall that $\delta<0$ in region
2):

\begin{equation}
\left(-\delta\right)<\frac{\lambda^{2}}{a_{p,0}}\left[\left(3-d\right)u_{m}-g_{m}\right]
\end{equation}

Therefore, we conclude that close enough to the tricritical point,
the nematic and magnetic transitions are simultaneous and first-order.
Note that the results resemble those of Ref. \onlinecite{Cano_PRB_2010},
although the models are somewhat different, as in that case the vestigial
Ising-nematic transition was not considered.

\section{Discussions}

\label{sec:Discussions}

\subsection{Dependence of the ordered moment on $T_{{\rm N}}$}

\label{subsec:Moment-on-Tn}

\begin{figure}[htb]
\includegraphics[width=0.95\linewidth]{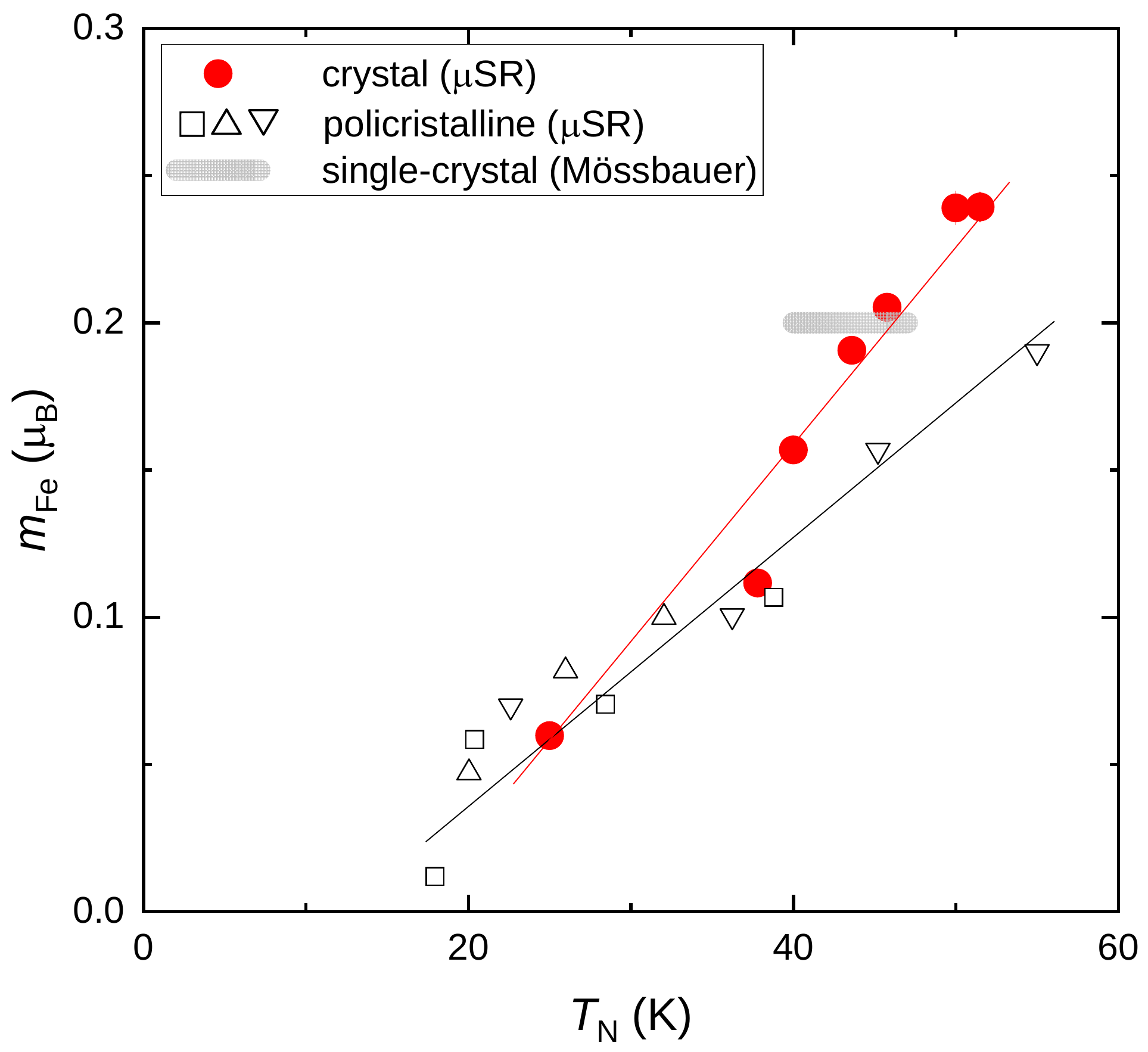} 
\caption{Value of the ordered magnetic moment on the Fe site $m_{{\rm Fe}}$
as a function of the magnetic ordering temperature $T_{{\rm N}}$.
The red circles correspond to the results of the present study on
101 oriented FeSe crystal. The open squares and the up and down triangles
are from $B_{{\rm int}}(T_{{\rm N}})$ $\mu$SR measurements of Bendele
\textit{et al.}\cite{Bendele_PRL_2010,Bendele_PRB_2012} on polycrystalline
FeSe samples synthesized by two different techniques. The grey stripe
correspond to the M\"{o}ssbauer results of Kothapalli \textit{et al.}\cite{Kothapalli_NatComm_2016}
on high-quality single crystalline FeSe sample. $T_{{\rm N}}$'s for
M\"{o}ssbauer data were obtained by fitting Eq.~\ref{eq:Power-Law} to
the temperature dependence of the hyperfine field $H_{{\rm pf}}$
at $p=2.5$ and 3.5~GPa {[}Fig.~2(b) in Ref.~\onlinecite{Kothapalli_NatComm_2016}{]}.
The solid lines are guides for the eyes.}
\label{fig:moment_vs_Tn}
\end{figure}

The internal field at the muon stopping site $B_{{\rm int}}$ is directly
proportional to the value of the ordered magnetic moment.\cite{Yaouanc_book_2011}
When the magnetic order is of stripe-type, $m_{{\rm Fe}}$ scales
with $B_{{\rm int}}$ as: $m_{{\rm Fe}}\simeq3.17\;\mu_{{\rm B}}/{\rm T}\;\times B_{{\rm int}}$
(see Ref.~\onlinecite{Khasanov_PRB_2017}). The value of the ordered
magnetic moment on the Fe site $m_{{\rm Fe}}$ as a function of the
magnetic ordering temperature $T_{{\rm N}}$ is shown in Fig.~\ref{fig:moment_vs_Tn}.
For comparison we have also included $m_{{\rm Fe}}$ values for various
FeSe samples (powders Refs.~\onlinecite{Bendele_PRL_2010, Bendele_PRB_2012}
and single crystal Ref.~\onlinecite{Kothapalli_NatComm_2016}) available
to date in the literature. Note that due to the small values of $m_{{\rm Fe}}$,
only $\mu$SR experiments permit a determination of $m_{{\rm Fe}}$ with
reliable accuracy. The M\"{o}ssbauer measurements of Kothapalli \textit{et
al.}\cite{Kothapalli_NatComm_2016} provide $m_{{\rm Fe}}\sim0.2\;\mu_{{\rm B}}$
for $p=2.5$ and 4.0~GPa, while the neutron experiments of Bendele
\textit{et al.}\cite{Bendele_PRB_2012} furnished just an upper estimate
of $m_{{\rm Fe}}\lesssim0.5-0.7\;\mu_{{\rm B}}$ at $p=4.4$~GPa.

Figure~\ref{fig:moment_vs_Tn} shows that $m_{{\rm Fe}}$ scales
linearly with $T_{{\rm N}}$. The highest value of the magnetic moment
$m_{{\rm Fe}}\simeq0.25\;\mu_{{\rm B}}$ correspond to the ordering
temperature $T_{{\rm N}}\simeq52$~K. The value of $m_{{\rm Fe}}$
for FeSe obtained in our study is one of the smallest among other
mother compounds of Fe-SC families (see \textit{e.g.} Ref.~\onlinecite{Dai_RMP_2015}
and references therein). It would be important to extend the experiments
up to higher pressures (at least up to 4-5~GPa), where according
to the results of Sun \textit{et al.}\cite{Sun_NatComm_2016} and B\"{o}hmer {\it et al.}\cite{Bohmer_Arxiv_2018} $T_{{\rm N}}$
reaches its maximum value. Unfortunately presently our pressure
cells do not allow to achieve pressures higher than $\sim2.7$~GPa.\cite{Khasanov_HPR_2016}

\subsection{Pressure dependence of the magnetic ordering temperature }

\label{subsec:Tn-on-p}

\begin{figure}[htb]
\includegraphics[width=0.9\linewidth]{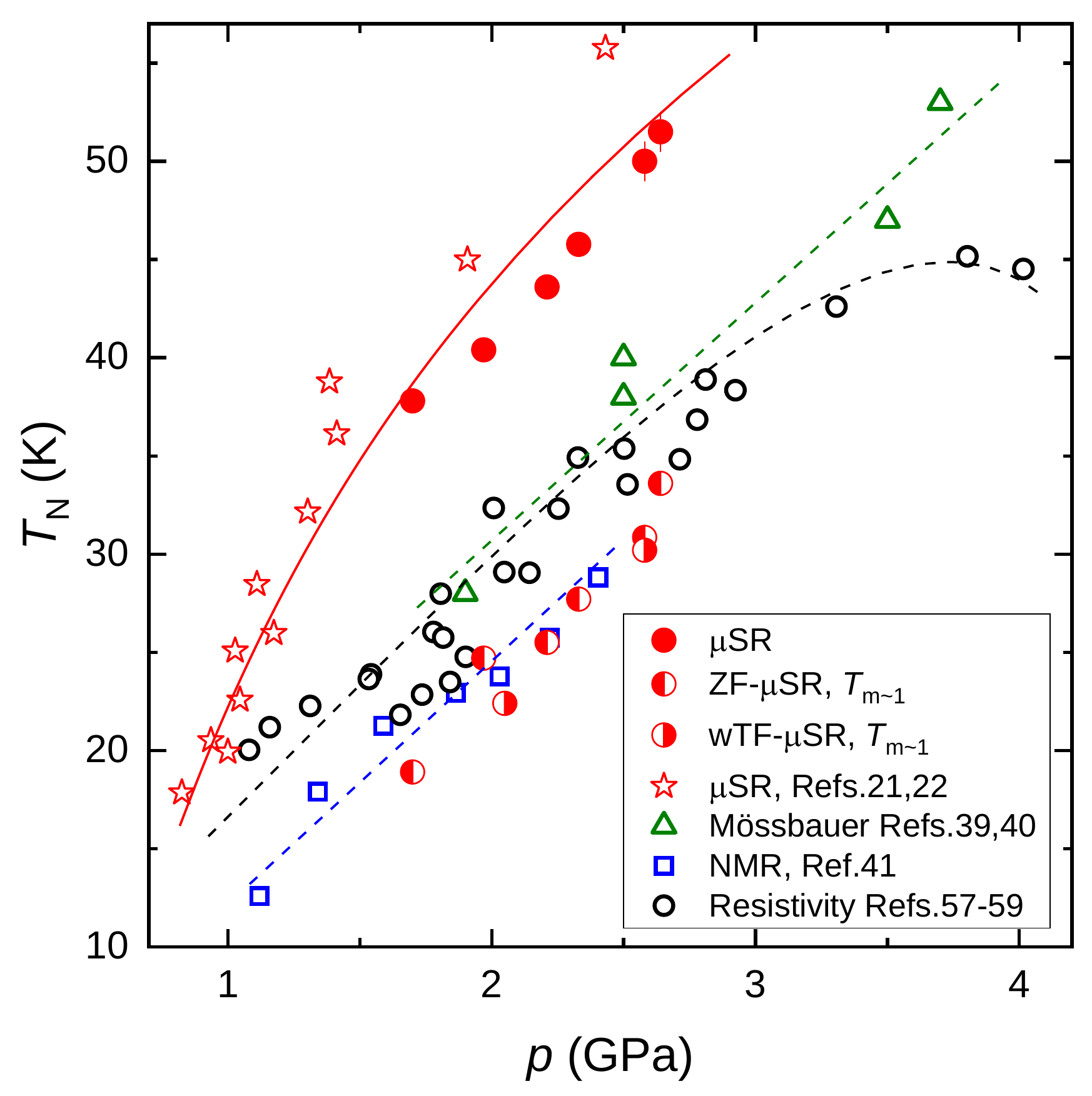} 
\caption{Dependence of the magnetic ordering temperature ($T_{{\rm N}}$) and
the temperature where the magnetic volume fraction starts to decrease
below unity ($T_{{\rm m\simeq1}}$) on applied pressure ($p$) in FeSe
obtained by different techniques. The red closed and semi open circles
correspond to the results of the present study on 101 oriented FeSe
crystal. The open stars are $\mu$SR data on FeSe polycrystaline samples.\cite{Bendele_PRL_2010,Bendele_PRB_2012}
Open squares and triangles are NMR and M\"{o}ssbauer data from Refs.~\onlinecite{Wang_PRL_2016}
and \onlinecite{Kothapalli_NatComm_2016,Bohmer_Arxiv_2018}, respectively. The open
circles correspond
to $T_{{\rm N}}$ obtained in resistivity experiments.\cite{Sun_NatComm_2016,Terashima_JPSJ_2015,Terashima_PRB_2016}
Lines are guides for the eyes. }
\label{fig:Tn_vs_p}
\end{figure}

The pressure induced magnetic transition in FeSe was previously studied
by $\mu$SR,\cite{Bendele_PRL_2010,Bendele_PRB_2012}, resistivity,\cite{Sun_NatComm_2016,Terashima_JPSJ_2015,Terashima_PRB_2016}
NMR,\cite{Wang_PRL_2016} and M\"{o}ssbauer experiments.\cite{Kothapalli_NatComm_2016,Bohmer_Arxiv_2018}
The dependencies of the magnetic ordering temperature on pressure
obtained in our studies and the above mentioned experiments are summarized
in Fig.~\ref{fig:Tn_vs_p}.

Figure~\ref{fig:Tn_vs_p} shows that there is a large spread in $T_{{\rm N}}$
values obtained by different techniques. The $\mu$SR experiments
on polycrystalline samples (Refs.~\onlinecite{Bendele_PRL_2010, Bendele_PRB_2012})
and crystal (present study) have found the highest $T_{{\rm N}}$
values. The NMR measurements, on the other hand, gave the lowest values
of $T_{{\rm N}}$.\cite{Wang_PRL_2016} $T_{{\rm N}}$ obtained from M\"{o}ssbauer and resistivity
experiments (Refs.~\onlinecite{Kothapalli_NatComm_2016, Bohmer_Arxiv_2018, Sun_NatComm_2016, Terashima_JPSJ_2015, Terashima_PRB_2016}) lie in between
the $\mu$SR and NMR data. Another big difference is the width of
the magnetic transition. Note that for this one can only compare
the $\mu$SR and NMR data. There is no criteria to estimate
the width of the magnetic transition from the resistivity and M\"{o}ssbauer data. In
$\mu$SR experiments the magnetic volume fraction $m$ decreases gradually
from its maximum value at low temperatures to an almost zero within a
rather broad temperature range [$\sim20-30$~K, see Figs.~\ref{fig:Bint_Fraction}~(c), (d)
and Fig.~\ref{fig:WTF}]. In NMR experiments, however, the magnetic transition,
which is associated with an abrupt change of the spectral weight,
has at most a 2-3~K transition width.\cite{Wang_PRL_2016}

One may determine the temperature $T_{{\rm m\sim1}}$ at which the
magnetic volume fraction $m$, as measured in ZF- and wTF-$\mu$SR
experiments, starts to deviate from unity [see Figs.~\ref{fig:Bint_Fraction}~(c),
(d) and Fig.~\ref{fig:WTF}]. The criteria for obtaining $T_{{\rm m\sim1}}$
is presented in Fig.~\ref{fig:Bint_Fraction}~(d). It is remarkable
that the $T_{{\rm m\sim1}}$ values plotted in Fig.~\ref{fig:Tn_vs_p}
\textit{coincide} with the $T_{{\rm N}}$ values observed by means
of NMR. We believe, therefore, that the spread of $T_{{\rm N}}$ shown
in Fig.~\ref{fig:Tn_vs_p} is caused by inhomogeneity of the samples
studied.

Indeed, the polycrystalline samples studied previously by $\mu$SR (Refs.~\onlinecite{Bendele_PRL_2010,
Bendele_PRB_2012}) were found to contain some amount of magnetic
impurities. This was confirmed by a series of powder neutron diffraction,
magnetization and ZF-$\mu$SR experiments.\cite{Khasanov_PRB_2008,Pomjakushina_PRB_2009}
The x-ray studies of a crystal, which was grown similarly to the
one used in the present study, reveal the presence of the hexagonal
impurity phase.\cite{Ma_SST_2014} The residual resistance ratio (RRR)
was estimated to be ${\rm RRR}\simeq6$. Our ambient pressure ZF-$\mu$SR
experiments (not shown) reveal an exponential character of the muon
polarization decay which might be explained by a static magnetic field
distribution caused by diluted and randomly oriented magnetic moments.\cite{Khasanov_PRB_2008}
In contrast, the sample studied by NMR in Ref.~\onlinecite{Wang_PRL_2016}
was supposed to be much more ``clean'' with ${\rm RRR}\simeq20$.

Cui {\it et al.}~\cite{Cui_Arxiv_2018} proposed
that disorder has a dramatic effect on nematicity, particularly near
a putative nematic quantum critical point, where finite-size droplets
can harbor long-range nematic order in the non-ordered side of the
phase transition. This leads to appearance of an {\it inhomogeneous} nematically-ordered
state developing up to higher values of the control parameter (doping or pressure) in comparison with that in a clean (free of impurity) system. Since the nematic order can enhance the magnetic ordering temperature, an inhomogeneous distribution of nematic transition temperatures could cause the observed decrease of the magnetic volume
fraction and the increase of $T_{\rm N}$.

At this point we would also note that the disorder in
FeSe samples has a strong influence not only on the magnetic, but also
on the superconducting properties.\cite{Rossler_Arxiv_2017}

\subsection{Comparison between the experiment and the theory}
\label{subsec:Consistency-Second-First}

The fact that the nematic transition temperature initially decreases
with pressure and then increases again once it meets the magnetic
transition line suggests that the nematic instability of unpressurized
FeSe has a different origin than the nematicity at higher pressures
(see also Refs. \onlinecite{Kothapalli_NatComm_2016, Bohmer_JPCM_2017, Bohmer_Arxiv_2018}). Since the
phase diagram of FeSe at high pressures is reminiscent of the usual iron-pnictide
phase diagram,\cite{Wang_PRL_2016,Kothapalli_NatComm_2016,Bohmer_Arxiv_2018} where nematicity is likely a vestigial phase of the
stripe magnetic order,\cite{Fernandes_PRL_2013} it is expected that the
origin of the nematic transition at low pressures involves a different
mechanism.

The results presented here of a magnetic tricritical point tuned by
pressure allow one to further test this hypothesis. The phenomenological
model presented in Sec.~\ref{sec:Theoretical-model} shows that the
emergence of this magnetic tricritical point is not only consistent,
but is generally expected if the nematic order parameter at
low pressures is different than the vestigial Ising-nematic state
arising at high pressures. The model shows that the tricritical
point at $p=p_{c}$ occurs slightly before the magnetic and nematic
transition lines meet at $p^{*}\gtrsim p_{c}$.

Indeed, our results for pressures up to $p\simeq2.33$~GPa show that
the magnetic order parameter ($B_{{\rm int}}\propto m_{{\rm Fe}}$)
decreases continuously down to zero as temperature increases, which is a clear indication of the second-order transition [see Fig.~\ref{fig:Bint_Fraction}~(a)].
For pressures in the range of $2.33<p<2.58$~GPa the magnetic transition
changes from second-order to first-order, indicating that the tricritical
point at $p_{c}$ is within this range. Note that the critical pressure
values obtained by other techniques are relatively close to the $p_{{\rm c}}$
region obtained in our study. The NMR experiments result in $p_{{\rm c}}\simeq2.2$~GPa,\cite{Wang_PRL_2016}
while the M\"{o}ssbauer data suggest $p_{{\rm c}}\gtrsim2.5$~GPa.\cite{Kothapalli_NatComm_2016}
Finally, for the highest pressures reached in the experiment (2.58
and 2.64~GPa), the magnetic order parameter drops abruptly thus demonstrating
that the magnetic transition becomes first-order [Fig.~\ref{fig:Bint_Fraction}~(b)].

Several theoretical models were proposed to explain the unusual
nematic order of unpressurized FeSe. \cite{Valenti_NatPhys_2015, Yu_PRL_2015,DHLee_NatPhys_2015,Kontani_PRX_2016,Fanfarillo_PRB_2018,Chubukov_PRX_2016}
While the phenomenological model discussed here does not allow us
to distinguish between these different scenarios, our results could be compared with the theory calculations reported in Ref.~\onlinecite{Chubukov_PRX_2016}.
There, the instabilities of the system towards superconductivity,
magnetism, and orbital order were treated on equal footing by a renormalization-group
analysis of a general microscopic itinerant model that explicitly
takes into account orbital degrees of freedom. The outcome depends on whether energy scale $T_{\rm ins}$ associated with the leading instability is smaller or larger than the Fermi energy $E_{\rm F}$. For $T_{\rm ins}>E_{\rm F}$, the leading instability is the Pomeranchuk one, followed by $s^{+-}$ superconductivity. For $T_{\rm ins}<E_{\rm F}$, the leading instability changes to the magnetic
one. Magnetic fluctuations that drive Ising-nematic order also favor a sign-changing $s^{+-}$ superconducting state. The link to the scenario
proposed in Ref. \onlinecite{Chubukov_PRX_2016} assumes that pressure shifts the balance between $E_{\rm F}$ and $T_{\rm ins}$ in favor
of magnetism (and vestigial Ising-nematicity) instead of the non-magnetically
ordered Pomeranchuk state. If pressure indeed enhances $E_{F}$, it
would offer an appealing microscopic mechanism for the phenomenological
model proposed in Sec.~\ref{sec:Theoretical-model}.

\section{Conclusions}

\label{sec:Conclusions}

We have performed a detailed study of the pressure-induced magnetic
order in the Fe-based binary pnictide superconductor FeSe. Zero-field
and weak transverse-field $\mu$SR experiments within the pressure
range of $1.7\lesssim p\lesssim2.64$~GPa on a FeSe crystal sample
are presented. Pressure induced magnetic order associated with
the appearance of the spontaneous muon-spin precession was detected.
The main experimental results are summarized as follows: (i) The value
of the ordered magnetic moment per Fe atom ($m_{{\rm Fe}}$) increases
linearly with increasing pressure, reaching $m_{{\rm Fe}}\simeq0.25$~$\mu_{{\rm B}}$
at $p\simeq2.6$~GPa. (ii) The magnetic order becomes more homogenous
at higher pressures. The distribution of magnetic moments decreases
from $\Delta m_{{\rm Fe}}/m_{{\rm Fe}}\simeq45$\% at $p=1.72$~GPa
to $\simeq19$\% at $p=2.64$~GPa.\cite{comment} (iii) The magnetic transition
changes from second- to first-order at a critical pressure $p_{c}\simeq2.4-2.5$~GPa.
(iv) Both $T_{{\rm N}}$ and $m_{{\rm Fe}}$ increases linearly with
increasing pressure thus suggesting that the stripe-type magnetic
order remains unchanged above and below the critical pressure $p_{{\rm c}}$.
(v) Comparison of the magnetic ordering temperature $T_{{\rm N}}$
with the results presented up to date in the literature suggests that
the onset of the magnetic transition in FeSe is determined by the
sample homogeneity. In homogenous FeSe samples the transition into
the magnetic state is sharp. In inhomogeneous samples the nematic
and magnetic orders may survive locally up to higher temperatures
than that expected for the homogeneous sample.

These observations and, in particular, the emergence
of a magnetic tricritical point are expected within a scenario where
the origin of the nematic transition at high pressures is similar
to other Fe-SC's, arising as a vestigial state from the stripe
magnetic order. On the other hand, at low pressures, nematicity likely
arises from a different mechanism. While this mechanism may still
involve magnetic fluctuations, as proposed for instance in Refs. \onlinecite{Valenti_NatPhys_2015,DHLee_NatPhys_2015,Chubukov_PRX_2016},
it remains distinct from the condensation of a composite magnetic
order expected near the onset of stripe SDW.

\section*{Acknowledgments}

This work was performed at the Swiss Muon Source (S$\mu$S), Paul
Scherrer Institute (PSI, Switzerland). RMF is supported by the US Department of Energy, Office
of Science, Basic Energy Sciences, under Award DE-SC0012336.
The work of GS is supported
by the Swiss National Science Foundation, grants $200021\_149486$
and $200021\_175935$. The work of XD, FZ and ZZ was supported by
the \char`\"{}National Key Research and Development Program of China
(Grant 2016YFA0300301)\char`\"{}, the \char`\"{}Strategic Priority
Research Program (B)\char`\"{} of the Chinese Academy of Sciences
(Grant XDB07020100) and the Natural Science Foundation of China (Grant
11574370).

\end{document}